# SUBMICRON MULTI-BUNCH BPM FOR CLIC*

H. Schmickler, L. Soby, CERN, Geneva, Switzerland
A. Lunin, N. Solyak, M. Wendt#, V. Yakovlev, Fermilab, Batavia, IL 60510, U.S.A

## Abstract

A common-mode free cavity BPM is currently under development at Fermilab within the ILC-CLIC collaboration. This monitor will be operated in a CLIC Main Linac multi-bunch regime, and needs to provide both, high spatial and time resolution. We present the design concept, numerical analysis, investigation on tolerances and error effects, as well as simulations on the signal response applying a multi-bunch stimulus.

## INTRODUCTION

The proposed CERN linear collider (CLIC) requires a very precise measurement of beam trajectory to preserve the low emittance when transporting the beam through the Main Linac [1]. An energy chirp within the bunch train will be applied to measure and minimize the dispersion effects, which require high resolution (in both, time and space) beam position monitors (BPM) along the beam-line. We propose a low-Q waveguide loaded $TM_{110}$ dipole mode cavity as BPM, which is complemented by a $TM_{010}$ monopole mode resonator of same resonant frequency for reference signal purposes. The design is based on a well known $TM_{110}$ selective mode coupling idea [2,3]

The BPM design process consists of several aspects:
- cavity spectrum calculations
- estimation of parasitic signals of monopole and quadruple modes,
- orthogonal ports cross coupling calculation
- and finally an analysis of the mechanical tolerances of the geometric structure.

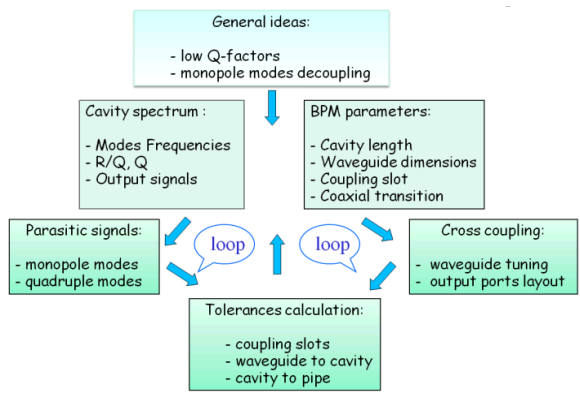

Figure 1: The BPM design process diagram.

___________________________________________
*Work supported by the Fermi National Accelerator laboratory, operated by Fermi Research Alliance LLC, under contract No. DE-AC02-07CH11359 with the US Department of Energy.
#manfred@fnal.gov

The results of each step depend on others and therefore the design process splits on several iterative loops (see Fig. 1).

The required design parameters of the BPM are given in the Table 1. Choosing a rather high operating frequency $n \, f_{bunch}$ has several advantages, e.g. most higher-order modes (HOM) are damped by the beam pipe cut-off frequency, and higher shunt impedances can be achieved (better sensitivity, higher resolution potential). However, as dipole mode and reference cavity operate at the same frequency, we have to ensure that they do not couple by evanescence fields leaking into the beam pipe

Table 1: CLIC Main Linac BPM specifications

| | |
|---|---|
| Nominal bunch charge [nC] | 0.6 |
| Bunch length (RMS) [µm] | 44 |
| Batch length [ns] | 156 |
| Bunch spacing [ns] | 0.5 |
| Beam pipe radius [mm] | 4 |
| BPM time resolution [ns] | <50 |
| BPM spatial resolution [nm] | <50 |
| BPM stability [nm] | <100 |
| BPM accuracy [µm] | <5 |
| BPM dynamic range [µm] | ±100 |
| BPM resonator frequency [GHz] | 14 |

With a time resolution of <50ns we will be able to acquire three beam position samples within the 156ns long bunch train (batch). Because of dynamic range limitations in the read-out system, the high 50nm spatial resolution can only be accomplished within a small range of ±100µm beam displacement from the BPM center, however a moderate resolution (few µm) will be achievable over the full aperture (±4mm).

## CAVITY BPM DESIGN

The proposed CLIC BPM consists of a cylindrical cavity loaded with four slot-coupled rectangular waveguides [4]. The schematic BPM view is shown in Figure 2. The off-axis beam passing the cavity induces two orthogonal dipole $TM_{110}$ modes with amplitudes proportional to the off-axis shift. A resonant cavity behaves like a damped oscillator with the EM- field decaying exponentially in time:

$$v(t) \propto v_0 \, e^{-t/\tau} \cos(\omega_0 t) \qquad (1)$$

where $\tau=2Q/\omega_0$ is resonator time constant, and Q is loaded quality factor. The required BPM time resolution (< 50ns) and dynamic range (±100μm) limits the maximum loaded quality factor of the cavity. The amplitude of the $TM_{110}$ dipole mode should decay $10^3$ times within 50ns in order to achieve the required resolution. Thus, the maximum loaded Q-factor is given by:

$$Q_{max} = \frac{t_{max}\omega_0}{2\ln(1000)} \approx 320 \qquad (2)$$

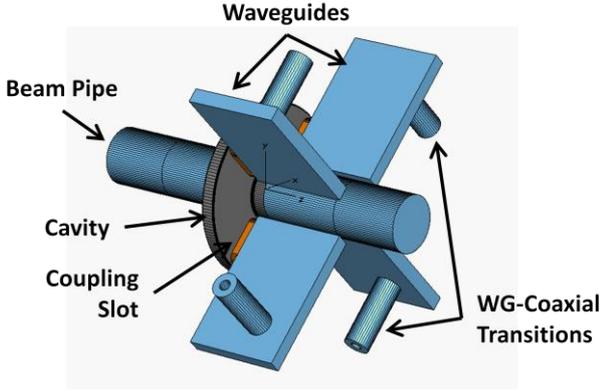

Figure 2: CLIC cavity BPM.

Keeping the cavity dimensions reasonable and applying the waveguide load impedance is not sufficient to achieve this rather low Q-value, thus the resonator has to be manufactured out of a more lossy material, i.e. stainless steel. The length of waveguide was optimized in order to eliminate any trapped resonances near working frequency 14GHz. We selected a 50Ω ultra-high vacuum microwave feed-through which is available by several vendors, but will need some custom modifications. The dimension of the feed-through pin was matched to the waveguide by a resonance antenna coupling method.

## BPM SPECTRUM CALCULATION

The BPM cavity, with its four slot-coupled waveguides, terminated by WG-coaxial transition ports, and beam pipe ports is a complex resonant system. A single beam bunch will excite most eigenmodes. In order to find a proper cavity radius and length we have to analyze each mode in the spectrum, characterizing resonant frequencies, Q-factors, $R/Q$ and output voltages. Such calculations were done using *ANSOFT* HFSS. However, only the $TM_{110}$ cavity dipole mode is of interest for our beam position measurement. Figure 3 illustrates the $TM_{110}$ mode coupling mechanism.

The other, unwanted modes, even though they have different resonance frequencies, will perturb the $TM_{110}$ dipole mode because their energy is spread over a wide range of frequencies, due to their limited Q-value (mode leakage). For each mode below the beam pipe cut-off frequency (~22GHz) we estimated the signal voltage (at t=0) as a sum from all coaxial ports:

$$v_0 = q\omega\sqrt{\frac{Z_0(R/Q)}{Q_{ext}}} \qquad (3)$$

where $q$ is the bunch charge, $Z_0$=50Ω is the characteristic impedance of the coaxial port, and $R/Q$ is the characteristic impedance of the cavity for a specific mode. At this point we found the initial cavity and coupling slots parameters enabling to obtain the required $Q_{ext}$ value. Finding the final BPM dimensions requires further tolerances analysis on the limitation of the BPM spatial resolution.

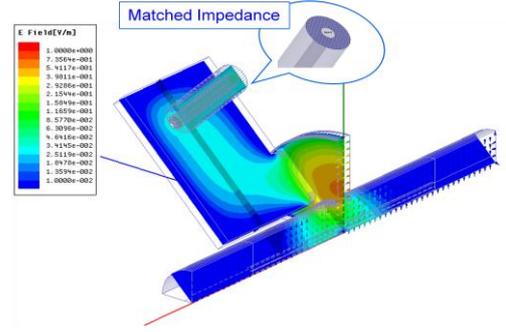

Figure 3: E-field of the $TM_{110}$ eigenmode.

The computed output voltage for each mode has to be multiplied by the expected rejection coefficients, i.e. frequency selectivity, phase filtering and multi-bunch rejection. The frequency filter rejection $KF_n$ is simply the normalized signal intensity $v_n$ of an unwanted mode $\omega_n$ leaking into the $TM_{110}$ dipole mode (at $f_{110}$):

$$KF_n = \frac{v_n(\omega_n)}{v_n(\omega_{110})} \qquad (4)$$

The proposed BPM has a two pair of outputs. Because of EM-field components exiting an opposite slots are shifted by 180 degree for dipole and monopole modes one can assumed that external symmetry rejection by an order of magnitude is possible by connecting a hybrid junction. In fact the phase of monopole output signal is undefined as it depends randomly on mechanical tolerances. Nevertheless, by summing the two opposite port signals we can benefit in additional ~20dB (0.1) phase filter rejection for the quadruple $TM_{210}$ mode.

In the practical application, the cavity BPM will not be excited by a single bunch, but by a train of typically $k$=312 bunches, spaced by 0.5ns. Therefore the output signal is a superposition of the single bunch responses of all modes, assuming the same intensity of all bunches. While $f_{110}$ is in phase with the bunch frequency $1/t_b$, resulting in a positive signal pile-up, the unwanted modes receive a "random" multi-bunch excitation, which leads to a rejection with respect to the beam position signal:

$$KM_n = \frac{\int_0^{t_{max}} \sum_1^k V_{110} e^{-\beta_{110}(t-t_bk)} \cos[\omega_{110}(t-t_bk)]}{\int_0^{t_{max}} \sum_1^k V_n e^{-\beta_n(t-t_bk)} \cos[\omega_n(t-t_bk)]} \qquad (5)$$

where $t_{max}$ is a required BPM time resolution.

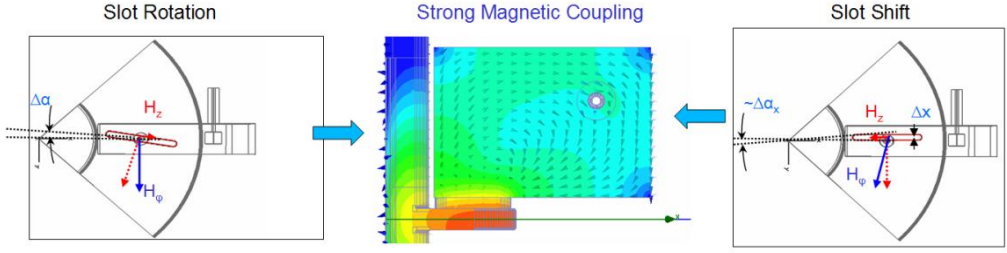

Figure 4: Investigation of tolerances on the cavity-waveguide coupling slot

## SPATIAL RESOLUTION LIMITS

The spatial resolution of the cavity BPM is defined as the smallest change of the beam position, which can be resolved. Beside the mode leakage effects, we also have to investigate consequences of practical imperfections, like mechanical tolerances from the manufacturing process, and other imperfections. It turns out, that the alignment of the coupling slot windows between cavity resonator and waveguides are most critical. In detailed EM-simulations we studied the effects of asymmetries due to shift, rotation and tilt of these slots, which causes additional leakage of the higher-order modes (see Figure 4). Also a shift between beam pipe and cavity centers will cause a degradation of the spatial resolution. In summary, limiting the slot shift to <5μm, and the slot rotation to <0.05deg, we can expect the cross coupling between horizontal and vertical plane to be >40dB, while achieving a ±100μm dynamic range at maximum resolution. This resolution is dominated by the $TM_{010}$ mode leakage, and is ~40nm for a single bunch (SB) excitation, and ~4nm for a multi-bunch (MB) beam batch (see Table 3). At large beam displacements >0.5mm, the $TM_{210}$ mode leakage dominates and limits the theoretical achievable resolution, however, in practice the limited dynamic range of the read-out receiver will further reduce the BPM resolution for this case.

Table 3: Limitations of the BPM resolution due to $TM_{010}$ & $TM_{210}$ mode leakage.

| Mode Type | Freq. [GHz] | $Q_{tot}$[1] | Beam shift [μm] | Output voltage[2] [mV] | BPM Resolution [nm] SB | BPM Resolution [nm] MB |
|---|---|---|---|---|---|---|
| $TM_{010}$ | 10.385 | 380 | 0 | <1 | 40 | 4 |
| $TM_{110}$ | 13.999 | 250 | 0.1 | 2.4 | - | - |
| $TM_{210}$ | 18.465 | 80 | 100 | <0.18 | 8 | 1 |
| $TM_{210}$ | 18.465 | 80 | 500 | <4 | 200 | 20 |

[1] – Stainless steel material was used.

[2] – RMS value of the sum signal of two opposite coaxial ports at the 14GHz operating frequency after all filters applied; signals are normalized to 1nC charge

The cross coupling between the two polarizations of the $TM_{110}$ mode also limits a dynamic range of the beam position measurement. The actual effect of cross coupling depends on amplitude and phase of reflected signals from the read-out electronics front-end, e.g. LLRF parts like hybrids or band-pass filters. For our estimation we assumed a worst case scenario, i.e. the reflected signals are in-phase and the SWR of the LLRF components is about -20dB. The required mechanical tolerances of a cavity with coupling slots are summarized in Table.4.

Table 4. Limitations of BPM resolution due to $TM_{110}$ modes cross coupling.

| Mechanical Tolerances[1,2] | Cross Coupling | | |
|---|---|---|---|
| | -40dB | -30dB | -20dB |
| Slot rotation [deg] | < 0.05 | < 0.2 | < 0.6 |
| Slot shift [μm] | < 5 | < 15 | < 40 |
| Max dynamic range [μm] | 100 | 25 | 10 |

[1] - In-phase signals reflection (worse case) is taken into account.
[2] - The reflection from LLRF part is assumed less than -20dB

## SUMMARY

A simple, straightforward design of a cavity BPM with high spatial (50nm) and temporal (50ns) resolution is proposed for the CLIC Main Linac. In depth EM-simulations and optimizations including the analysis of mechanical tolerances were performed in order to prove the BPM design .parameters.